\documentclass{aa}  
\usepackage{graphicx}
\usepackage{natbib}
\bibpunct{(}{)}{;}{a}{}{,}
\usepackage{txfonts}
\begin{document}
   \title{The planet-hosting subdwarf B star \object{V\,391 Pegasi} 
          is a hybrid pulsator%
          \thanks{Based on observations collected at the Centro 
       Astron\'omico Hispano Alem\'an (CAHA) at Calar Alto,
       operated jointly by the Max-Planck Institut f\"ur
       Astronomie and the Instituto de Astrof\'{\i}sica de Andaluc\'{\i}a
       (CSIC).}
	  }

   \author{R. Lutz\inst{1,4}
          \and
          S. Schuh\inst{1}
	  \and
	  R. Silvotti\inst{2}
	  \and
          S. Bernabei\inst{3}
	  \and
	  S. Dreizler\inst{1}
	  \and
	  T. Stahn\inst{4}
	  \and
	  S. D. H\"ugelmeyer\inst{1}
          }


   \institute{Institut f\"ur Astrophysik, Universit\"at G\"ottingen,
              Friedrich-Hund-Platz~1, 37077 G\"ottingen, Germany\\
              \email{rlutz@astro.physik.uni-goettingen.de}
         \and INAF - Osservatorio Astronomico di Capodimonte, via Moiariello 16, 
              80131 Napoli, Italy
         \and INAF - Osservatorio Astronomico di Bologna, via Ranzani 1, 40127 Bologna, Italy
	 \and Max-Planck-Institut f\"ur Sonnensystemforschung,
              Max-Planck-Stra\ss e 2, 37191 Katlenburg-Lindau, Germany
             }

   \date{Received; accepted}

 
  \abstract
   {A noticeable fraction of subdwarf B stars
     shows either short-period ($p$-mode) or long-period ($g$-mode)
     luminosity variations, with two objects so far known to exhibit
     hybrid behaviour, i.e.\ showing both types of modes at the same
     time. The pulsating subdwarf B star \object{V\,391 Pegasi} (or
     \object{HS\,2201+2610}), which is close to the two known hybrid
     pulsators in the $\log g$\,--\,$T_{\rm eff}$ plane, has recently been
     discovered to host a planetary companion.}
   {In order to learn more about the planetary companion and its
     possible influence on the evolution of its host star (subdwarf B star
     formation is still not well understood), an accurate
     characterisation of the host star is required. As part of an ongoing
     effort to significantly improve the asteroseismic
     characterisation of the host star, we investigate the
     low-frequency behaviour of \object{HS\,2201+2610}.}
   {We obtained rapid high signal-to-noise photometric CCD (B-filter) and PMT
     (clear-filter) data at 2\,m-class telescopes and carried out a careful 
     frequency analysis of the light curves.}
   {In addition to the previously known short-period luminosity
     variations in the range 342\,s\,$-$\,367\,s, we find a long-period
     variation with a period of 54\,min and an amplitude of 0.15 per
     cent. This can most plausibly be identified with a $g$-mode pulsation, so 
     that \object{HS\,2201+2610} is a new addition to the short
     list of hybrid sdB pulsators.}
   {Along with the previously known pulsating subdwarf B stars
     \object{HS\,0702+6043} and \object{Balloon\,090100001} showing
     hybrid behaviour, the new hybrid \object{HS\,2201+2610} is the
     third member of this class. This important property of
     \object{HS\,2201+2610} can lead to a better characterisation of this
     planet-hosting star, helping the characterisation of its planetary
     companion as well. Current pulsation models cannot yet
     reproduce hybrid sdBV stars particularly well and improved pulsation
     models for this object have to include the hybrid behaviour.
   } 


   \maketitle
%

\section{Introduction}
\label{sec:introduction}
Subdwarf B stars (sdBs) are evolved objects with masses
around half a solar mass and are thought to be core helium burning. Having
lost most of their original hydrogen envelope in earlier stages of
their evolution, the remaining H-rich envelope is too thin to sustain
hydrogen shell burning.  The sdBs populate the extreme horizontal
branch (EHB) at effective temperatures of 20\,000 to 40\,000~K
and surface gravities $\log (g/\rm{cm\,s^{-2}})$ between 5.0 and
6.2. Instead of evolving towards the asymptotic giant branch, the
sdBs are predicted to follow tracks leading directly towards the white
dwarf region after leaving the extreme horizontal branch.
A fraction of the sdB stars shows pulsations: the first
pulsating sdB star was discovered in 1997. Variable subdwarf B stars 
(sdBV stars) can be divided into the classes of rapid \emph{$p$-mode pulsators}
(sdBV$_\textrm{r}$) and slow \emph{$g$-mode pulsators} (sdBV$_\textrm{s}$), 
with two objects known so far to belong to both classes simultaneously 
(\emph{hybrid pulsators}, sdBV$_\textrm{rs}$). 

The $p$-mode pulsators show low amplitudes (few ten mmag) and short periods
(few minutes) at higher temperatures. The acoustic pressure waves are of low
degree and low radial order. In contrast, the $g$-mode pulsators have even
lower amplitudes (few mmag) and longer periods (30 to 90~min) at lower
temperatures. The class prototypes are
\object{EC\,14026-2647} \citep{1997MNRAS.285..640K} and \object{PG\,1716+426}
\citep{2003ApJ...583L..31G}, respectively. Both mode types are thought to be
driven by the same $\kappa$-mechanism due to an opacity bump of iron
and nickel \citep{1997ApJ...483L.123C,
  2003ApJ...597..518F, 2006MNRAS.372L..48J}. Hybrid sdB pulsators show $p$- and
$g$-modes simultaneously and are of particular interest since the two mode 
types probe different regions within the star. In the 
$\log g$\,--\,$T_{\rm eff}$ diagram, the hybrids are located at the interface 
of the $p$-mode and $g$-mode instability regions. The known hybrids are
\object{HS\,0702+6043} \citep{2005ASPC..334..530S, 2006A&A...445L..31S,
  2008ASPC..392..339L} and \object{Balloon\,090100001}
\citep{2005A&A...438..257O, 2005MNRAS.360..737B}. 

This paper presents the third member of this group:
\object{HS\,2201+2610} is a 14.3 B-magnitude star 
having an effective temperature of 29\,300$\pm$500~K and a surface gravity
$\log g$ of 5.40$\pm$0.10 in cgs units \citep{2001A&A...368..175O}. Based on 
these spectroscopic parameters, which are very similar to those of 
\object{HS\,0702+6043} and \object{Balloon\,090100001}, \object{HS\,2201+2610} 
was a natural further target to search for simultaneous $g$-modes in a known 
$p$-mode
\begin{figure}[ht!]
 \resizebox{\hsize}{!}{\includegraphics[angle=90]{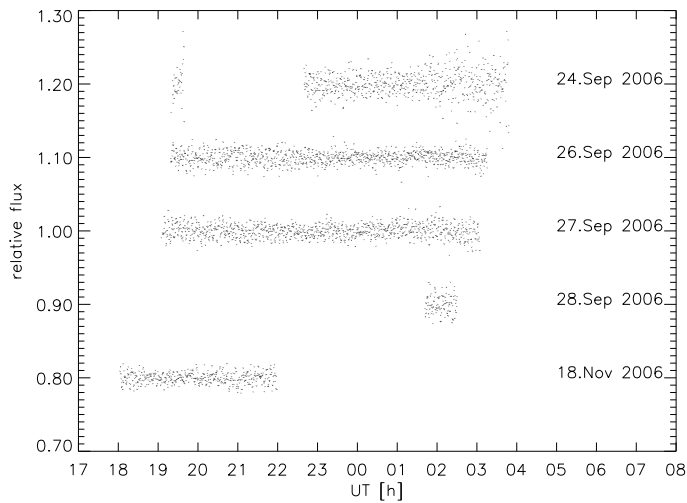}}
 \caption{Light curves from the September/November 2006 runs at Calar
 Alto. The relative flux is set to 1.0 for the 27.\,Sep 2006 light curve and
 previous (following) nights are shifted upwards (downwards) by 0.1 units.}
 \label{fig:lc}
\end{figure}
pulsator after the discovery of the first hybrids. The high
signal-to-noise observations initiated to check for hybrid behaviour
in \object{HS\,2201+2610} fitted in well with our ongoing long-term
monitoring in order to search for secular changes in the
pulsation periods indicative of slow evolutionary changes in the star
\citep{2002A&A...389..180S}, which has recently resulted in the
surprising detection of a planetary companion around
\object{HS\,2201+2610} (better known as \object{V\,391 Pegasi b},
\citealt{2007Natur.449..189S}).
In fact, from a few of the better-quality data sets obtained within
this overall observational effort up to the end of 2005, we already
suspected the presence of additional power in the low-frequency domain of
\object{HS\,2201+2610}. 

In Section~\ref{sec:observations} we present the photometric
data obtained specifically to investigate the gravity mode frequency
range, as well as suitable confirmation data obtained as part of our
long-term monitoring. The data reduction and frequency analysis
are described in Section~\ref{sec:results}, and as a result of our investigation we will
present \object{HS\,2201+2610} as a new member of the group of hybrid
pulsating sdB stars. We discuss the implications of this discovery for
the ongoing asteroseismic characterisation of \object{HS\,2201+2610}, in
the context of our efforts to constrain the mass of its planetary
companion, in Section~\ref{sec:discussion}. Finally, we present a concluding
outlook in Section~\ref{sec:conclusions}.

\section{Observations}
\label{sec:observations}

The light curves (see Fig.\,\ref{fig:lc}) were obtained at Calar Alto 
Observatory in September and November 2006. The instrumental setup was chosen 
to consist of the Calar Alto Faint Object Spectrograph (CAFOS) focal reducer 
attached to the 2.2~m telescope. A 2k\,$\times$\,2k CCD chip recorded the 
science frames through a Johnson B filter with a 2\,$\times$\,2 binning and an 
exposure time of ten seconds with an average cycle time of 25.5 seconds. In
order to sample the short-period $p$-modes as well as to detect very
low-amplitude $g$-modes, one has to find a good compromise with the exposure
time. Ten seconds turned out to be sufficient to resolve the $p$-mode cycles as
well as provide a S/N good enough to resolve low-amplitude signals. We 
have a total of about 26 hours of photometric data.

 \begin{figure}[ht!]
 \resizebox{\hsize}{!}{\includegraphics[angle=90]{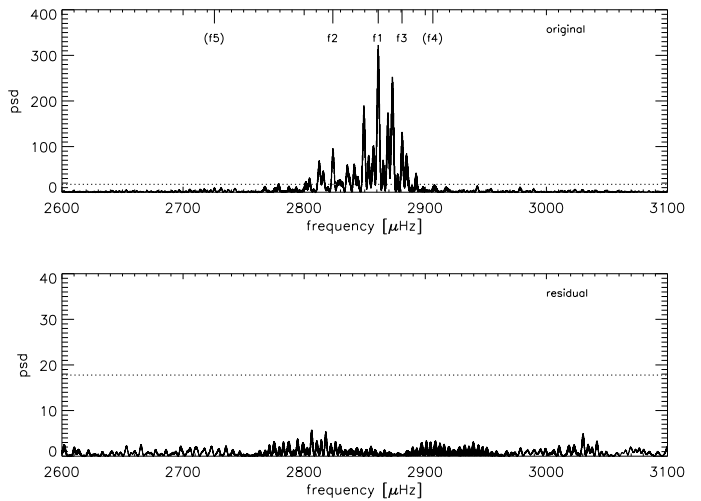}}
 \caption{Lomb Scargle Periodograms (LSPs) of the $p$-mode regime. The top 
     panel shows the complete Calar Alto data. The bottom panel displays the 
     LSP after simultaneous subtraction of five frequencies. Horizontal dotted 
     lines indicate 4~$\sigma$ confidence levels and the marks correspond to 
     the IDs in Table \ref{table:frequencyanalysis}.}
 \label{fig:pw_p}
 \end{figure}

The data reduction was carried out using the Time Resolved Imaging Photometry
Package (TRIPP, see \citealt{2003BaltA..12..167S}), which is designed to handle
the aperture photometry of large CCD sets. This includes bias-,
flatfield- and background corrections and finally yields extracted light
curves in the form of relative flux (in comparison to four stable reference
stars) as a function of geocentric Julian date. Additionally, a barycentric
time correction was applied. Differential extinction effects on a 
night-by-night basis were corrected using second-order-polynomials.

In addition to our Calar Alto data from 2006, one of us (R.~S.) provided older
data from August/September 2003, which were part of the Whole Earth Telescope
extended coverage 23 (WET Xcov23). The data taken at Loiano 
Observatory (LOI, 1.5~m telescope and 3 channel photometer) and Observatoire 
de Haute-Provence (OHP, 1.9~m telescope and 3 channel photometer) were 
provided as extracted light curves (total length of 28 hours) and we used them 
to cross-check our Calar Alto results.

\section{Results}
\label{sec:results}
We carried out a frequency analysis on both data sets with a
program called \texttt{sinfit}, which is part of TRIPP. This 
includes least-squares
non-linear sinusoidal fits to the light curves and calculation of the 
corresponding periods, amplitudes and phases. The amplitudes are given in 
milli modulation intensity units (mmi). In this context 10~mmi means that 
the luminosity has changed 
by one per cent with respect to the mean brightness level. We used the Scargle algorithm 
\citep{1982ApJ...263..835S} to calculate Lomb-Scargle-Periodograms 
(LSPs) in the form of power spectral density (psd) as a function of frequency, 
together with confidence levels, which are based on false alarm probabilities 
(faps). See Section 4 in \citet{2002A&A...386..249D} for a more detailed
description. In our case, we derived the faps from 158000 white noise
simulations. This yields a confidence level 
of 4~$\sigma$, judging a signal above this level to be real with a 
probability of 99.99366\%.
%
 
   \begin{figure*}
   \centering
   \includegraphics[angle=90,width=\textwidth]{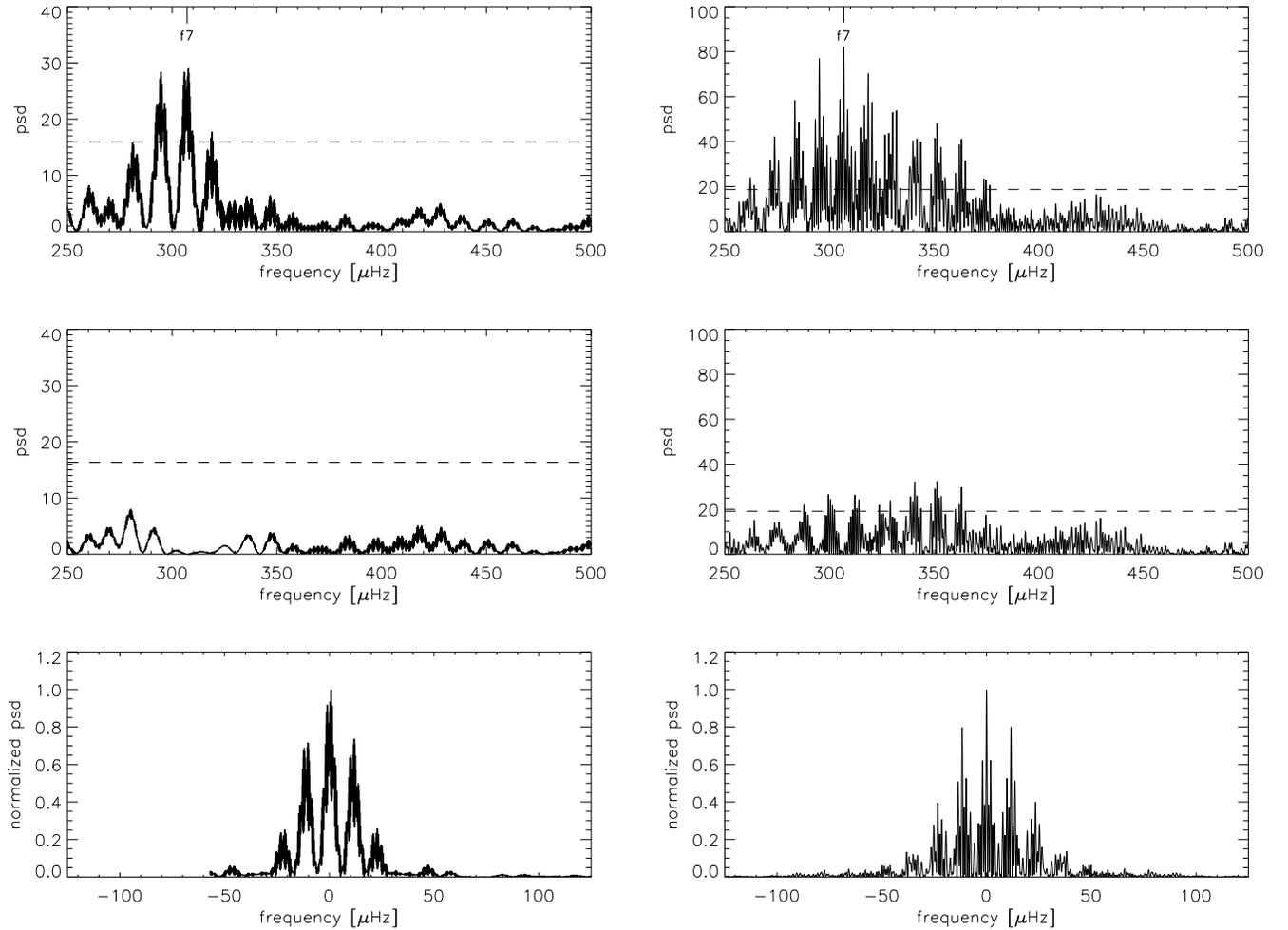}
   \caption{Results of our frequency analysis in the $g$-mode
     regime. \textit{Left}: September/November 2006 Calar Alto data. The 
     top panel is a Lomb Scargle Periodogram of the three best nights
     of the full light curve where five 
     frequencies have already been subtracted (f1 to f5, see 
     Table \ref{table:frequencyanalysis}). Fitting a long-periodic luminosity 
     variation (f7) to this subset of the data results in the residual LSP displayed in the 
     middle. The window function is shown in the bottom panel. 
     \textit{Right}: August/September 2003 OHP/LOI data. The top panel
     displays the Lomb Scargle Periodogram of the light 
     curve after pre-whitening of three frequencies (f1 to f3, see Table
     \ref{table:frequencyanalysis}). The residual LSP after subtraction of the
     long-period variation f7 is shown in the middle and the window
     function is illustrated in the bottom panel. Horizontal dashed lines 
     indicate 4~$\sigma$ confidence levels.} 
   \label{fig:calaralto_vs_ohploi}
   \end{figure*}

Table \ref{table:frequencyanalysis} summarizes the results of our frequency 
analyses of both data sets. We can clearly recover the three strongest 
modes found in an extensive run published by \citet{2002A&A...389..180S}, here 
labeled with f1 to f3. These frequencies are assigned to pressure driven 
modes. The original and residual LSPs of our Calar Alto data in the $p$-mode 
regime are displayed in Fig.\,\ref{fig:pw_p}. In addition, we find a 
long-periodic luminosity variation (f7, shown in 
Fig.\,\ref{fig:calaralto_vs_ohploi}) of about 54 minutes. 
%
\begin{table*}
\caption{Results of our frequency analysis of
  \object{HS\,2201+2610}. We refer to the pulsation frequency identifications
  obtained by
  \citet{2002A&A...389..180S} and also list the mode degree (\textit{l})
  identification suggested there in Fig.\,9.
  \textit{Calar Alto data:} We estimate the errors in the frequency to be 
  0.2~$\mu$Hz (FWHM of the central
  peak in the window function) 
  and having a detection limit of 1.1~mmi, which was calculated as four times 
  the mean noise level in the residual DFT. 
  We used (f4) and (f5) in the prewhitening, because low-amplitude
  signals at these frequency ranges were  
  also found in other runs \citep{2002A&A...389..180S}. 
  \textit{OHP/LOI data:} We calculate a
  frequency resolution of 0.54~$\mu$Hz (FWHM of the central peak in the window 
  function) and a detection limit of 0.97~mmi (four times the mean noise level 
  in the residual DFT). From the WET Xcov23 runs, the OHP data from 
  August 22/23 and the LOI data from August 29/31 and September 3/4 have been 
  included.
  \textit{Mean solution:} The frequencies of the ''mean'' solution
  refer to BJD~2452788, which corresponds to the middle of the 7-year
  long data set in \citet{2007Natur.449..189S}. Estimating the formal frequency
  resolution of the 7-year ''mean'' solution yields 0.0045\,$\mu$Hz. 
}
\label{table:frequencyanalysis}      
\centering
\begin{tabular}{cllr@{}lr@{}lr@{}lr@{}lr@{}lr@{}lr@{}l}
\hline\hline
\multicolumn{5}{p{43mm}}{}&\multicolumn{4}{p{40mm}}{Calar Alto Data (2006)}&
\multicolumn{4}{p{40mm}}{OHP/LOI Data (2003)}&\multicolumn{4}{p{40mm}}{Mean
  solution (2000\,--\,2007)}\\
\hline    
ID & type &$l$&
\multicolumn{2}{l}{Period}&\multicolumn{2}{l}{Frequency}&\multicolumn{2}{l}{Amplitude}&
\multicolumn{2}{l}{Frequency}&\multicolumn{2}{l}{Amplitude}&
\multicolumn{2}{l}{Frequency}&\multicolumn{2}{l}{Amplitude}\\
   &      &   &
\multicolumn{2}{l}{[s]}&\multicolumn{2}{l}{[$\mu$Hz]}&\multicolumn{2}{l}{[mmi]}&
\multicolumn{2}{l}{[$\mu$Hz]}&\multicolumn{2}{l}{[mmi]}    &
\multicolumn{2}{l}{[$\mu$Hz]}&\multicolumn{2}{l}{[mmi]}    \\
\hline                    
f1  & $p$ &0& 349.&5 & 2861.&2 & 6.&6 & 2860.&9  &12.&3 & 2860.&9387 & 10.&4\\
f2  & $p$ &1& 354.&2 & 2823.&7 & 3.&3 & 2824.&1  & 3.&7 & 2824.&0965 &  4.&2\\
f3  & $p$ &1& 347.&1 & 2880.&8 & 1.&5 & 2880.&7  & 1.&5 & 2880.&6842 &  0.&7\\
 f4 & $p$ &2&     &  &      &  &   &  &      &   &   &  & 2921.&8463 &  0.&6\\
($\approx$ f4 alias)&  & & 344.&1 & 2906.&3 & 0.&4 &      &   &   &  &      &      &    & \\
(f5)& $p$ &2& \\ 
(f5 alias)& & & 366.&9 & 2725.&9 & 0.&8 &      &   &   &  &      &      &    & \\
(f6)    & $p$ & &     &  &      &  &   &  &      &   &   &  &(2882.&0039 &  0.&5)\\
f7  & $g$ & &3255.&6 &  307.&2 & 1.&5 &  306.&8  & 1.&5 &      &      &    & \\
\hline                  
\end{tabular}
\end{table*}

As mentioned before, we looked at suitable WET Xcov23 data with the focus on 
the $g$-mode domain to confirm our detection of a long-periodic luminosity
variation found in the Calar Alto data (left plots in 
Fig.\,\ref{fig:calaralto_vs_ohploi}). The periodogram of these additional
data is shown in the right plots of Fig.\,\ref{fig:calaralto_vs_ohploi}. 
The OHP/LOI results (also in Table \ref{table:frequencyanalysis}) clearly
confirm the detection of a low frequency oscillation in \object{HS\,2201+2610}.

As can be seen in the middle right plot in 
Fig.\,\ref{fig:calaralto_vs_ohploi}, 
there is still further residual power above 4~$\sigma$ in the OHP/LOI data LSP 
after subtraction of the long-period variation f7. However, this is not seen
in the Calar Alto data (which were taken three years later) so that further
high-quality data are required to test the persistence of these residual peaks.

\section{Discussion}
\label{sec:discussion}
We identify the long-period low-amplitude luminosity variation in 
\object{HS\,2201+2610}
with a gravity driven mode. The presence of both types 
of modes within \object{HS\,2201+2610} would therefore define this object to 
be a member of the hybrid sdB pulsator group, beside the previously known 
hybrids \object{HS\,0702+6043} and \object{Balloon\,090100001}. 
Other interpretations for the long-period variation can be excluded with a high
probability with similar arguments as for \object{HS\,0702+6043}
\citep{2006A&A...445L..31S}. 
The same long-period variation, in
frequency as well as amplitude, is present in data sets three years 
apart. This rules out sky or transparency variations, as well
as rotating star spots. The fast rotation rate that would be required for the latter
interpretation is further unlikely since rotation periods of single
sdB stars are typically much longer. 
Rotational splitting (for $l>0$ modes) has not yet been detected in
\object{HS\,2201+2610}, but the split components often have lower
amplitudes and might be below our detection limit. 
With the light-travel time effect used in the detection of the planet
candidate around \object{HS\,2201+2610} in its wide orbit, we would
not be sensitive to a further hypothetical companion in a very close
orbit with a 54\,min period. A binary origin of the long-period
variation is unlikely for the following reasons.
A 54\,min orbital period would be at the extreme low limit of the observed and
theoretical orbital period distributions (see e.g.\ Fig.\,2 in
\citealt{2006BaltA..15..187M}). Additionally, a low-mass main-sequence
  (M-dwarf) companion as the source of the 54\,min signal can be excluded by
  geometrical considerations (Roche-Limits vs.\ separation in Keplerian
  orbits). The same is true for a sdB+WD configuration in a 54\,min orbit down
to very low WD-masses (below 0.2\,M$_{\sun}$). In the case of deformation of the sdB
by a hypothetical WD companion, causing ellipsoidal variation, the underlying
orbital period would be 108\,min, but we would then expect a signal at least ten
times stronger down to low inclinations, dropping to the observed very
low amplitude at only about 10\degr. A WD companion as the source of the
signal is therefore also very unlikely.
For less massive and hence harder to detect companions it would be
increasingly hard to prevent mass transfer in the system.
Binary orbital motion as the cause of the long-period variation will
immediately and conclusively be ruled out if its multi-periodicity
is confirmed.
\par
The period and amplitude fall within the typical range for
$g$-mode sdB pulsations, strongly suggesting
that the variation is in fact due to a $g$-mode. 
Finally, in comparison to the location of the other two known hybrid
sdB pulsators in the $\log g$\,--\,$T_{\rm eff}$ diagram of sdBV stars
(see e.g.\ Fig.\,1 in \citealt{2008ASPC..392..339L}), the very
similar parameters of \object{HS\,2201+2610} again imply that hybrid
behaviour may plausibly be expected. All three objects are located at
the interface between the empirical $p$-mode and $g$-mode 
instability regions. Hybrid pulsators have also been shown to exist in
the partially overlapping regions of the $\beta$\,Cep / [SPB] stars, which resemble
the $p$ / $g$-mode sdB pulsators in many ways
\citep[e.g.][]{2004MNRAS.347..454H,2005MNRAS.360..619J,2006MNRAS.365..327H,2007arXiv0711.3179P,2007arXiv0711.2530P}.
Hybrid behaviour is also observed in 
the less closely related $\delta$\,Sct / $\gamma$\,Dor stars
\citep[e.g.][]{2005AJ....129.2026H}.
\par
From a theoretical point of view, modeling $p$-mode sdB pulsators has been 
very successful. \citet{2006ESASP.624E..32F} report on the determination of 
the basic structural parameters of twelve $p$-mode sdB pulsators. 
Regarding the $g$-mode sdB pulsators there are still issues. 
The ''second-generation models'' presented by
\citet{1997ApJ...483L.123C} and \citet{2003ApJ...597..518F} diverge by
some 4000~K from the observations, i.e.\ they fall short of the
empirical blue edge of the $g$-mode pulsators. A solution to this
problem has been suggested by
\citet{2006MNRAS.372L..48J,2007MNRAS.378..379J}: In their models,
using Opacity Project (OP) instead of OPAL opacity tables and including
\element[][]{Ni} in addition to \element[][]{Fe} to the opacity
profile shifts the theoretical blue edge close to the empirical
one. Both aspects of this opacity problem are, in close analogy, also
relevant for the $\beta$\,Cep and [SPB] pulsators \citep{2007MNRAS.375L..21M}. In fact,
the hybrid objects (both in the main-sequence as well as in the
extreme horizontal branch pulsators groups) in particular 
represent excellent test cases for the adequateness of the
opacities used, and should provide strong constraints.
From previous experience, it appears to be clear that radiative
levitation and self-consistent diffusion of the
\element[][]{Fe}/\element[][]{Ni} group elements need to be included
in the EHB models. It will be interesting to see if ''third-generation
models'' incorporating both radiative levitation/diffusion and improved 
opacities
will finally deliver an adequate description of the hybrid sdB
pulsators while still reproducing the behaviour of the $p$-mode
pulsators as consistently as before, and at the same time correctly
describing the extent of the empirical $g$-mode instability region.


\section{Concluding outlook}
\label{sec:conclusions}
As recently published by \citet{2007Natur.449..189S}, \object{HS\,2201+2610}
has a planetary companion which was detected using asteroseismological
techniques. Long-term monitoring of the strongest $p$-mode
pulsations made it possible to derive secular variations of the two main pulsation
frequencies, leading to an evolutionary timescale of around $5-8 \cdot 10^{6}$
years, which matches theoretical evolutionary models of extreme horizontal 
branch stars. Periodic residuals in the $O-C$ diagram (observed $-$
calculated) analysis, attributable to light-travel time effects, revealed the
presence of an $m\sin{i}=3.2\,\textrm{M}_{\textrm{Jup}}$ companion around
\object{HS\,2201+2610} in a 3.2\,yr orbit. 
We are in the process of continuing our regular long term photometric
monitoring of \object{HS\,2201+2610} beyond the current coverage of
two orbit cycles.
In order to further characterise the companion, especially to
determine its mass, it is necessary to constrain the orbital inclination 
of the system and to derive a more accurate asteroseimic
mass of the host star.
\par
The mass of the host star is in principle accessible through
asteroseismology, and \citet{2002A&A...389..180S} already presented one
possible solution. The determination of the fundamental parameters of
the star has however only weakly been constrained in that work due to
the small number of modes detected.
Now that we have added \object{HS\,2201+2610} to the list of hybrid
sdB pulsators, this finding might provide the additional constraints
necessary for a reliable and accurate asteroseismic characterisation. At
this point, we emphasize again the current need to improve on the model
calculations first. For the individual object \object{HS\,2201+2610},
it will be a special challenge to additionally drive $g$-modes in an
improved version of one of the possible models found by
\citet{2002A&A...389..180S}.

\begin{acknowledgements}
The Loiano and OHP data were taken during the Whole Earth Telescope extended
coverage run 23; we thank the WET Xcov23 team:
R.~Janulis, J.-E.~Solheim, R.~{\O}stensen, T.~D.~Oswalt,
I.~Bruni, R.~Gualandi, A.~Bonanno, G.~Vauclair, M.~Reed,
C.-W.~Chen, E.~Leibowitz, M.~Paparo, A.~Baran, S.~Charpinet,
N.~Dolez, S.~Kawaler, D.~Kurtz, P.~Moskalik, R.~Riddle, S.~Zola.
%
%
Part of the observations at Calar Alto were carried out in service
mode by U.~Thiele and A.~Guijarro.
The observations at Calar Alto were supported through Deut\-sche
For\-schungs\-ge\-mein\-schaft (DFG) travel grant \mbox{SCHU 2249/3-1}.
\end{acknowledgements}


\end{document}